\documentclass[journal]{IEEEtran}

\usepackage{amsmath}
\usepackage{bm}
\usepackage[colorlinks=true,bookmarks=false,citecolor=blue,urlcolor=blue]{hyperref} 
		\usepackage{subfig}
		\usepackage{graphicx}	
		\usepackage{cite}
\usepackage{epstopdf}	
\usepackage{multirow}

\begin{document}

\title{Nonlinear Propagation in Multimode and Multicore Fibers: Generalization of the Manakov Equations}

\author{Sami Mumtaz, Ren\'e-Jean Essiambre and Govind P. Agrawal, Fellow IEEE%
\thanks{Sami Mumtaz and Govind P. Agrawal are with The Institute of Optics, University of Rochester, Rochester, NY,14627,USA}
\thanks{ Ren\'e-Jean Essiambre is with Bell Laboratories, Alcatel-Lucent, 791 Holmdel, NJ, 07733, USA}%
\thanks{Manuscript received January xx, 2012; revised March xx, 2012. First published May xx, 2012.}
\thanks{This work was funded by Alcatel-Lucent in the framework of GreenTouch.}}

\markboth{Nonlinear propagation in multimode fibers}
{Shell \MakeLowercase{\textit{et al.}}: Bare Demo of IEEEtran.cls for Journals}

\maketitle

\begin{abstract}
This paper starts by an investigation of nonlinear transmission in space-division multiplexed (SDM) systems using multimode fibers exhibiting a rapidly varying birefringence. A primary objective is to generalize the Manakov equations, well known in the case of single-mode fibers. We first investigate a reference case where linear coupling among the spatial modes of the fiber is  weak and after averaging over birefringence fluctuations, we obtain  new Manakov equations for multimode fibers.  Such an averaging reduces the number of intermodal nonlinear terms drastically since all four-wave-mixing terms average out. Cross-phase modulation terms still affect  multimode transmission but their effectiveness is reduced.
We then verify the accuracy of our new Manakov equations by transmitting multiple 114-Gb/s bit streams in the PDM-QPSK format over different modes of a multimode fiber and comparing the numerical results with those obtained by solving the full stochastic equation. The agreement is excellent in all cases studied. A great benefit of the new equations is to reduce  the computation time by a factor of 10 or more. Another important feature observed is that birefringence fluctuations improve system performance by reducing the impact of fiber nonlinearities. The extent of improvement depends on the fiber design and how many spatial modes are used for SDM transmission. We investigated both  step-index and graded-index multimode fibers and considered up to three spatial modes in each case.
In this paper, we also consider multicore fibers and show that the Manakov equations remain valid as long as the length scale of birefringence fluctuations is about 100 times shorter than that of linear coupling. 
Finally multimode fibers with strong random coupling among all spatial modes are considered.  Linear coupling  is modeled using the random matrix theory approach. We derive new Manakov equations for multimode fibers in that regime and show that such fibers can perform better than single-modes fiber for large number of propagating spatial modes.
\end{abstract}


\IEEEpeerreviewmaketitle


\vspace{-4mm}
\section{Introduction}

Recent work on the optical fiber capacity limit \cite{Essiambre_JLT, ess11a ,mec12} has indicated that the continued use of single-mode fibers for modern telecommunication systems may not be able to support the growing data-traffic demand in the near future \cite{Tkach10,ess12b}. Although more single-mode fibers can be deployed as a short-term solution, it is important to look for alternative solutions for the next generation of optical transmission systems. The impressive bit-rate increase of the last two decades was made possible by exploiting diverse properties of the electromagnetic field through wavelength-division multiplexing (WDM) in combination with phase modulation formats and polarization-division multiplexing (PDM). 
Recently, multimode fibers have become the focus of attention \cite{RyfOFC11},\cite{Ip11} as they permit simultaneous propagation of multiple spatial modes. Multiplexing signals using different spatial modes of an optical fiber is referred as space-division multiplexing (SDM). Recent work on multimode fiber has focused on demonstrating the feasibility of the SDM technique and how digital signal processing (DSP) can help to compensate linear distortions of the signal \cite{RyfJLT12},\cite{ALUfr_ECOC}. However, nonlinear propagation effects  are the principal limitation of optical fiber communication systems and their understanding in the case of multimode fibers is still very limited compared to the case of single-mode fibers.

A set of nonlinear propagation equations has been derived for multimode fibers  in \cite{Poletti} by expanding the optical field in terms of vectorial fiber modes. It is difficult to use these equations in their most general form to study  birefringence effects. In the case of silica fibers used for telecommunication systems, we can simplify these equations by making use of the so-called weakly-guiding approximation. In this paper we derive in section~\ref{sec_2} a new set of multimode nonlinear equations, which represents the two polarization components of each spatial mode through a Jones vector and includes linear coupling between spatial modes. We then use them to study the effects of fiber's random modal birefringence and linear mode coupling using random matrices in two important cases of practical interest that we refer to as the weak and strong coupling regimes. In the weak-coupling regime, linear coupling among distinct spatial modes is weak compared to the birefringence-induced coupling between the two polarization components of the same spatial mode. In the strong-coupling regime, both types of coupling occur in a random fashion with the same order of magnitude.  In practice, certain spatial modes of multimode fibers may be weakly coupled whereas others, may be more strongly coupled. For instance, it was observed in \cite{RyfJLT12} that the LP11a spatial mode is strongly coupled with  LP11b but weakly coupled with LP01. The strong and weak spatial mode coupling regimes studied in this work represent two extreme cases for practical systems.  Multimode fibers always have, in general, some degree of coupling between any pair of spatial modes. Important factors impacting how much linear coupling there is between spatial modes are: fiber design and fabrication, cabling, bending and twisting. 

It is well known in the case of single-mode fibers that rapidly varying birefringence results in an averaging effect that reduces the strength of  nonlinearities: mathematically, this averaging is described by the Manakov equations \cite{Mana74,Meny96,Marc97}. Its use also reduces the computation time considerably when the nonlinear propagation is performed numerically. In general, the standard Manakov equations cannot be used in the case of multimode fibers since they only describe how birefringence impacts \emph{intramodal} nonlinear effects and do not include \emph{intermodal} nonlinear effects. In section \ref{sec_3}, we derive a new set of Manakov equations for nonlinear propagation in multimode fibers  in the regime of weak coupling between spatial modes, following the ideas in \cite{Meny96} and averaging over all possible polarization states. In section \ref{sec_4}, the impact of rapidly varying birefringence on a specific SDM transmission system is studied through numerical simulations. We show that the new set of Manakov equations successfully predict the effects of random modal birefringence, while also reducing the computational complexity of the problem. In section \ref{sec_5}, we apply the Manakov equations to the case of a multicore fiber whose cores are linearly coupled with a constant coupling coefficient. We show that Manakov equations remain valid in the weak coupling regime as long as the length scale of the birefringence fluctuations is 100 times shorter than that of linear coupling.

Finally, we consider in section \ref{sec_6} a case with large random coupling between spatial modes, referred to as the strong-coupling regime. Although current transmission experiments over multimode fibers are not fully in the strong-coupling regime \cite{RyfOFC11}, this case has attracted theoretical attention to obtain an understanding of stochastic effects such as random differential group delays among various modes \cite{HoJLT11} or random mode-dependent loss within the multimode fiber \cite{Ho11}. Recently Manakov equation have been generalized for multimode fibers in the high coupling regime \cite{Meco12}. Following a different approach based on random matrix theory, we obtain similar results within the weakly guiding approximation. A comparison of the weak- and strong-coupling regimes is also presented in this paper.

\section{Nonlinear propagation in multimode fibers} \label{sec_2}

We simplify the analysis in \cite{Poletti} by making the weakly guiding approximation, known to be quite accurate for silica fibers with a relatively low difference in the refractive indices between core and cladding. In this approximation, the modal fields are linearly polarized in the transverse plane, as the field components along the direction of propagation ($\rm z$ axis) are assumed to be negligible. Moreover, the spatial modes associated with the two polarization components  are assumed to have the same spatial distribution. With these simplifications, we write the total electric field as a sum over the $M$ distinct spatial modes (we use the same definition of the fiber  modes as in \cite{RyfJLT12}) of the fiber in the spectral domain:
\begin{equation}
    \widetilde{\mathbf{E}}({ x},{y},{z},\omega)= \sum_m^M e^{\imath{\bm \beta}_{m}(\omega){z}}\widetilde{\mathbf{A}}_m({z},\omega) \mathrm{F}_m({x},{y})/\sqrt{{\rm N}_{m}}, \label{eq_1}
\end{equation}
where $\widetilde{\mathbf{A}}_m(z,\omega)= [\widetilde{\mathrm{A}}_{m{\rm x}}(z,\omega)~\widetilde{\mathrm{A}}_{m{\rm y}}(z,\omega)]^{\rm T}$ is the Fourier transform of the time-domain slowly varying field envelope of the $m$th mode represented in the Jones vector notation. The superscript $^{\rm T}$ denotes the transpose operation and the tilde denotes a frequency-domain variable. It includes the slowly varying amplitudes of both polarization components for the spatial mode $m$ with the spatial distribution $\mathrm{F}_m({x},{y})$ and the propagation constant ${\bm \beta}_{m}(\omega)$, expressed  in the form of a diagonal matrix to account for  fiber birefringence, i.e., ${\bm \beta}_{m}={\rm diag}[\beta_{m{\rm x}}~\beta_{m{\rm y}}]$; $e^{\imath{\bm \beta}_{m}z}$ corresponds to the usual definition of an exponential matrix.
Note that $\mathrm{F}_m({x},{y})$ is real in the weakly guiding approximation, and the phase of the optical envelope is contained in the complex field $\widetilde{\mathbf{A}}_m$. Finally, the normalization constant ${\rm N}_{m}$ in (1) represents the power carried by the $m$th mode. It can be expressed as ${\rm N}_{m}= \frac{1}{2}\,\epsilon_0 \,\bar{n}_{\rm eff} \,c \,{\rm I}_m$, where ${\rm I}_m= \bar{n}_{m}/\bar{n}_{\rm eff}  \iint \mathrm{F}_m^2(x,y) dx\,dy$, $\epsilon_0$ is the vacuum permittivity,  $\bar{n}_{\rm eff}$ is the fiber effective index of the fundamental mode and $\bar{n}_{m}$ the fiber effective index of the mode $m$.

Each frequency component of the optical field satisfies the frequency-domain Maxwell equations,
\begin{equation}
    \nabla^2\widetilde{\mathbf{E}}(\omega) + {n}_{}^2(x,y,z) \frac{\omega^2}{c^2} \widetilde{\mathbf{E}}(\omega)= -\omega^2\mu_0\widetilde{\mathbf{P}}^{\rm NL}(\omega) \label{eq_2},
\end{equation}
where ${n}(x,y,z)$ is the refractive index distribution within the fiber whose $z$-dependence accounts for  small perturbations along the propagation distance, i.e., $n^2(x,y,z)=n_0^2(x,y)+\Delta n^2(x,y,z)$ \cite{Ghatak}. $\widetilde{\mathbf{P}}^{\rm NL}(\omega)$ is the Fourier transform of the third-order nonlinear response \cite{Boyd}
\begin{equation}
    {\mathbf{P}}^{\rm NL}(t)= \frac{\epsilon_0}{4}\chi^{(3)}\left([ \mathbf{E}^{\rm T} \mathbf{E}] \mathbf{E}^* + 2[\mathbf{E}^{\rm H}\mathbf{E}] \mathbf{E}\right).
\end{equation}
Here $^{*}$ denotes the complex conjugate and the superscript $^{\rm H}$ denotes the Hermitian conjugate. $\chi^{(3)}$ is the instantaneous nonlinear response of  silica.

We assume that the spatial distribution of a mode is not perturbed by  nonlinearities or by small variations of the refractive index along the propagation distance $z$. In that case, $\mathrm{F}_m(x,y)$ satisfies the eigenvalue equation
\begin{equation}
    \nabla^2{\mathrm{F}}_m(x,y) + \left(\frac{\omega^2}{c^2}{n}_{0}^2(x,y) -{\beta}^2_{mi}\right) {\mathrm{F}_m}(x,y)= 0, \label{eq_4}
  \end{equation}
where the subscript $i={\rm x,y}$ refers to the two orthogonal polarizations of the $m$th spatial mode. In fibers with perfect cylindrically symmetric refractive index profile $n_{0}(x,y)$, ${ \beta}_{m{\rm x}}= {\beta}_{m{\rm y}}$. In practice, because of imperfections, fibers exhibit some polarization birefringence and ${ \beta}_{m{\rm x}}\neq{ \beta}_{m{\rm y}}$. In the remainder of this section we consider a sufficiently short fiber section such that the birefringence remains constant along the fiber length. The case of accumulation of randomly oriented birefringent fibers segments is treated in subsequent sections. Note that the various solutions of (\ref{eq_4}) are chosen to be orthogonal to each other and are normalized such that
\begin{equation}
    \iint {\mathrm{F}}_m(x,y){\mathrm{F}}_p(x,y){d}x\,{d}y={\rm I}_m \delta_{mp},
\end{equation}
$\delta_{mp}$ being the Kronecker function.

To isolate the evolution of a specific spatial mode, say the $p$th mode, we use their orthogonal nature. We substitute (\ref{eq_1}) in (\ref{eq_2}), multiply both sides with ${\mathrm{F}}_p$, and integrate over the transverse $x$-$y$ plane. The resulting equation is then converted to the time domain by following a standard procedure and expanding ${\bm \beta}_p(\omega)$ in a Taylor series around the carrier frequency $\omega_0$ \cite{Agrawal}. The final result can be written in the form
\begin{eqnarray}
    \frac{\partial\mathbf{A}_p}{\partial{z}} 
   &=&~ \imath ( {\bm \beta}_{0p}\!-\!{ \beta}_r  ) \mathbf{A}_p-( {\bm \beta}_{1p}\!-\! \frac{1}{v_{g_r}}  )\frac{\partial \mathbf{A}_p}{\partial{ t}} - \imath\frac{{\beta}_{2p}}{2}\frac{\partial^2\mathbf{A}_p}{\partial{ t}^2} \nonumber\\
    &&~+~\imath\sum_{lmn} f_{lmnp}\frac{\gamma}{3}\left[(\mathbf{A}_l^{\rm T}
    \mathbf{A}_m^{}) \mathbf{A}_n^{*}+2(\mathbf{A}_l^{\rm H} \mathbf{A}_m^{}) \mathbf{A}_n^{}\right] \nonumber\\
    && ~+~\imath\sum_m q_{mp} \mathbf{A}_m, \label{eq_NLSE}
\end{eqnarray}
where $\mathbf{A}_p(z, t)$ is the time-domain slowly varying envelope of the $p$th mode expressed in a reference moving frame at the  group velocity $v_{g_{r}}$. $ {\beta}_{r} $ is an arbitrary reference propagation constant for all the spatial modes and   $ {\beta}_{0p}\!=\!\beta_p(\omega_0) $, ${\bm\beta}_{1p}\!=\!\partial {\bm \beta}_{p}/\partial \omega|_{\omega_0}$,
 ${\beta}_{2p}\!=\!\partial^2 {\bm \beta}_{p}/\partial \omega^2|_{\omega_0}$ are respectively the propagation constant, inverse group velocity and  group-velocity dispersion (GVD) of the $p$th spatial mode. We assume here that the two polarization components of a spatial mode may have different group velocities but have the same GVD.

The nonlinear parameter in (\ref{eq_NLSE}) is defined in a similar fashion as in single-mode fibers, by $\gamma=3k_0 \chi^{(3)}/(4 \epsilon_0 c n_{\rm eff}^2 A_{\rm eff})$, where $A_{\rm eff}$ is the effective area of the fundamental mode. The linear and nonlinear coupling  among spatial modes is governed, respectively, by
\begin{eqnarray}
    {q}_{mp}(z)&=&\frac{k_0}{2n_{\rm eff}({\rm I}_m{\rm I}_p)^{1/2}}
    \iint\Delta n^2(x,y,z)\mathrm{F}_m^{}\mathrm{F}_p\,dx\,dy,\\
    f_{lmnp}&=&\frac{A_{\rm eff}}{({\rm I}_l{\rm I}_m{\rm I}_n{\rm I}_p)^{1/2}}\iint\mathrm{F}_l\mathrm{F}_m\mathrm{F}_n\mathrm{F}_p\,dx\,dy.
\end{eqnarray}
Note that linear coupling terms appear as a result of the perturbation of the refractive index $\Delta n^2(x,y,z)$. In the case of an ideal fiber, $\Delta n^2(x,y,z)=0$, no linear coupling occurs between the spatial modes.

Equation~(\ref{eq_NLSE}) governs propagation of arbitrarily polarized light in the spatial mode $p$ within the Jones-matrix formalism. It includes all third-order nonlinear effects (both intramodal and intermodal kinds) as well as dispersive and birefringence effects. Fiber losses can also be included by adding the term $-(\alpha_p/2)A_p$ on its right side, where the loss parameter $\alpha_p$ can be different for different modes.


\section{Manakov Equation in Weak-Coupling Regime} \label{sec_3}

Equation~(\ref{eq_NLSE}) assumes  constant birefringence along the fiber length. In practice, owing to fiber imperfections such as a nonuniform core whose shape varies along the fiber, birefringence varies randomly and rapidly on a length scale that is short compared to the effective lengths associated with the GVD and various nonlinearities. This feature can be implemented numerically by rotating the principal axes of the fiber periodically after a distance shorter than the length scale of birefringence fluctuations. However, a numerical solution of Eq.~\ref{eq_NLSE} then requires a relatively small step size and is quite time-consuming.

In the case of single-mode fibers, the problem has been solved by adopting the well-known Manakov equation \cite{Meny96}\cite{Marc97}. The idea is that, as a rapidly varying birefringence changes continually the state of polarization (SOP) of propagating light in a random fashion, one can average the propagation equation itself over all polarization states. We can follow the same procedure for multimode fibers by assuming that the SOP of each spatial mode evolves, randomly and \emph{independently}, i.e., we assume that the SOP of each mode evolves in an independent fashion. This can be justified by noting that, as the spatial distributions of the modes are different, the influence of  local stress and fiber imperfections may also be different from one spatial mode to another.

A rapidly varying fiber birefringence can be tracked using the transformation
\begin{equation}
    \mathbf{A}_p(z)=\mathbf{R}_p(z)\mathbf{\bar{A}}_p(z)\label{eq_localaxis}
\end{equation}
where $\mathbf{R}_p(z)$, a unitary matrix belonging to the group U(2)  is a Jones matrix of the form
\begin{eqnarray}
    \mathbf{R}_p(z)=\left[\begin{matrix}
    r_{11p}^{}~ & ~r_{12p}^{} \\
    r_{21p}^{}~ & ~r_{22p}^{}
    \end{matrix}\right]\label{eq_matrot}\\\nonumber\\
   \mathbf{R}_p^H \mathbf{R}_p=\mathbf{I}_2
   \end{eqnarray}
where $\mathbf{I}_2$ is the $2\times 2$ identity matrix and  $r_{ijp}$ are random variables. An important issue is how to construct $\mathbf{R}_p$ to ensure that the SOP probability for the $p$th mode is uniformly distributed over all possible SOPs. We make use of the following procedure applicable to
the general case of n$\times $n matrices  \cite{Mezz07}: pick a matrix $\mathbf{M}$ whose all elements are normally distributed (with zero mean) and apply a QR decomposition, $\mathbf{M}=\mathbf{RT}$, such that $\mathbf{T}$ is an upper triangular matrix with positive diagonal entries. Then, $\mathbf{R}$ is a Haar matrix whose elements are uniformly distributed over U(n).

The use of (\ref{eq_localaxis}) in (\ref{eq_NLSE}), leads to the following equation for $\bar{\mathbf{A}}_p$:
\begin{eqnarray}
    \frac{\partial\bar{\mathbf{A}}_p}{\partial z} & =&
    \imath\, {\bm\delta}{\bm \beta}_{0p} \,\bar{\mathbf{A}}_p - {\bm \delta} {\bm \beta}_{1p}\,\frac{\partial \bar{\mathbf{A}}_p}{\partial t}
    -\imath\frac{{{\bm \beta}}_{2p}}{2}\frac{\partial^2 \bar{\mathbf{A}}_p}{\partial t^2}  \nonumber\\
    &&{+} \imath \sum_{lmn}f_{lmnp}\frac{\gamma}{3}
    \left( \left[\bar{\mathbf{A}}_l^{\rm T} ~\mathbf{R}_l^{\rm T}\mathbf{R}_m^{}\bar{\mathbf{A}}_m\right] \mathbf{R}_p^{\rm H}
    \mathbf{R}_n^{*}\bar{\mathbf{A}}_n^*\right.\nonumber\\
    &&\left.~~~~~+2
    \left[\bar{\mathbf{A}}_l^{\rm H} \mathbf{R}_l^{\rm H}\mathbf{R}_m^{}\bar{\mathbf{A}}_m\right]\mathbf{R}_p^{\rm H}
    \mathbf{R}_n^{} \bar{\mathbf{A}}_n \right) \nonumber\\
    && + \imath\sum_m \bar{\mathbf{{q}}}_{mp}\bar{\mathbf{A}}_m
\label{eq_main}
\end{eqnarray}
where
\begin{eqnarray}
   {\bm \delta}{\bm \beta}_{0p}&=& \mathbf{R}_p^{\rm H}\, ( {\bm \beta}_{0p}\!-\!{ \beta}_r  ) \, \mathbf{R}_p-\imath\mathbf{R}_p^{\rm H}\frac{\partial \mathbf{R}_p}{\partial z},\\
   {\bm \delta}{\bm \beta}_{1p}&=& \mathbf{R}_p^{\rm H}\,( {\bm \beta}_{1p}\!-\! \frac{1}{v_{g_r}}  )\, \mathbf{R}_p,\\
    \bar{\mathbf{{q}}}_{mp}&=&{ q}_{mp}\mathbf{R}_p^{\rm H}\mathbf{R}_m.\label{eq_17}
\end{eqnarray}

Equation (\ref{eq_main}) is stochastic because $\mathbf{R}_m(z)$ is a random matrix that changes along the fiber length on a length scale associated with birefringence fluctuations. As a result, the birefringence parameters appearing there vary randomly. In addition, the intramodal and intermodal nonlinear couplings, as well as the linear coupling, also become random.

In the following analysis, we average the propagation equation (\ref{eq_main}) over all possible realizations of the matrix $\mathbf{R}_m$. Averaging over birefringence amounts to assuming that other phenomena that produce $z$-dependent variations, in particular the linear coupling between spatial modes ($q_{mp}(z)$  in~(\ref{eq_17})), occur over a much longer length scale in comparison to the length scale of  polarization fluctuations within each spatial mode in the weak-coupling regime considered here.

Let us focus first on the nonlinear terms in (\ref{eq_main}) by writing them as
\begin{eqnarray}\label{gpa1}
\mathcal{N}_p&=&~ \imath \sum_{lmn}f_{plmn}\frac{\gamma}{3}
\left( \left[\bar{\mathbf{A}}_l^{\rm T} ~\mathbf{R}_l^{\rm T}\mathbf{R}_m^{}\bar{\mathbf{A}}_m\right]
\mathbf{R}_p^{\rm H} \mathbf{R}_n^{*}\bar{\mathbf{A}}_n^*\right.\nonumber\\
&&{+}\left.2
 \left[\bar{\mathbf{A}}_l^{\rm H} \mathbf{R}_l^{\rm H}\mathbf{R}_m^{}\bar{\mathbf{A}}_m\right]
\mathbf{R}_p^{\rm H} \mathbf{R}_n^{} \bar{\mathbf{A}}_n \right).
\end{eqnarray}
The matrix multiplications appearing there can be expressed as
\begin{equation}\label{gpa2}
    \mathbf{R}_l^{\rm T}\mathbf{R}_m^{}=\left[\begin{matrix}
    {\hat a}_{lm}^{} ~&~  {\hat b}_{lm}^{}\\  {\hat b}^{}_{ml}~&~ {\hat c}^{}_{lm}
 \end{matrix}\right] \quad
    \mathbf{R}_l^{\rm H}\mathbf{R}_m^{}=\left[\begin{matrix}
    a_{lm}~&~b_{lm} \\ b_{ml}^*~&~c^{}_{lm},
 \end{matrix}\right]
\end{equation}
where we have defined the following quantities:
\begin{eqnarray}
    a_{lm}&=& r_{11l}^* \,r_{11m}^{}+ r_{21l}^*\, r_{21m}^{}\\
    b_{lm}&=&r_{11l}^* \,r_{12m}^{}+ r_{21l}^*\, r_{22m}^{}\\
    c_{lm}&=&r_{21l}^* \,r_{21m}^{}+ r_{22l}^*\, r_{22m}^{}\\
    {\hat a}_{lm}&=&r_{11l}^{} \,r_{11m}^{}+ r_{21l}^{}\, r_{21m}^{}\\
    {\hat b}_{lm}&=&r_{11l}^{} \,r_{12m}^{}+ r_{21l}^{}\, r_{22m}^{}\\
    {\hat c}_{lm}&=&r_{21l}^{} \,r_{21m}^{}+ r_{22l}^{}\, r_{22m}^{}.
\end{eqnarray}

Using (\ref{gpa2}) in (\ref{gpa1}), the nonlinear terms for the $\rm x$-polarized $p$th mode become:
\begin{eqnarray}
    \mathcal{N}_{p{\rm x}}&=& \imath \sum_{lmn}f_{plmn}\frac{\gamma}{3}{\Big (}\nonumber\\
    &&2\Big[a^{}_{lm}a^{}_{pn}\bar{\mathrm{A}}_{l{\rm x}}^*\bar{\mathrm{A}}^{}_{m{\rm x}}\bar{\mathrm{A}}^{}_{n{\rm x}}   +  b^{*}_{ml}a^{}_{pn} \bar{\mathrm{A}}_{l{\rm y}}^*\bar{\mathrm{A}}^{}_{m{\rm x}}\bar{\mathrm{A}}^{}_{n{\rm x}} \nonumber\\
    &&~~{+}b^{}_{lm}a^{}_{pn}  \bar{\mathrm{A}}_{l{\rm x}}^*\bar{\mathrm{A}}^{}_{m{\rm y}}\bar{\mathrm{A}}^{}_{n{\rm x}} + ~  c^{}_{lm}a^{}_{pn}  \bar{\mathrm{A}}_{l{\rm y}}^*\bar{\mathrm{A}}^{}_{m{\rm y}}\bar{\mathrm{A}}^{}_{n{\rm x}}\nonumber\\
    &&~~{+}a^{}_{lm}b_{pn}^* \bar{\mathrm{A}}_{l{\rm x}}^*\bar{\mathrm{A}}^{}_{m{\rm x}}\bar{\mathrm{A}}^{}_{n{\rm y}}  +   b^{*}_{ml}b_{pn}^* \bar{\mathrm{A}}_{l{\rm y}}^*\bar{\mathrm{A}}^{}_{m{\rm x}}\bar{\mathrm{A}}^{}_{n{\rm y}} \nonumber\\
    &&~+b^{}_{lm}b^{}_{pn}  \bar{\mathrm{A}}_{l{\rm x}}^*\bar{\mathrm{A}}^{}_{m{\rm y}}\bar{\mathrm{A}}^{}_{n{\rm y}}   +   c^{}_{lm}a_{pn}^* \bar{\mathrm{A}}_{l{\rm y}}^*\bar{\mathrm{A}}^{}_{m{\rm y}}\bar{\mathrm{A}}^{}_{n{\rm y}} \Big]   \nonumber \\
    &&+\left[  {\hat a}^{}_{lm}\hat{a}_{pn}^* \bar{\mathrm{A}}^{}_{l{\rm x}}\bar{\mathrm{A}}^{}_{m{\rm x}}\bar{\mathrm{A}}_{n{\rm x}} ^*     +  \hat{b}^{}_{ml}\hat{a}_{pn}^*  \bar{\mathrm{A}}^{}_{l{\rm y}}\bar{\mathrm{A}}^{}_{m{\rm x}}\bar{\mathrm{A}}_{n{\rm x}} ^*   \right.  \nonumber \\
    &&~~ +  \hat{ b}^{}_{lm}\hat{a}_{pn}^* \bar{\mathrm{A}}^{}_{l{\rm x}}\bar{\mathrm{A}}^{}_{m{\rm y}}\bar{\mathrm{A}}_{n{\rm x}} ^*      +   ~\hat{c}^{}_{lm}\hat{a}_{pn}^* \bar{\mathrm{A}}^{}_{l{\rm y}}\bar{\mathrm{A}}^{}_{m{\rm y}}\bar{\mathrm{A}}_{n{\rm x}} ^*  \nonumber\\
    &&~~  +  \hat{a}^{}_{lm}\hat{b}_{pn}^* \bar{\mathrm{A}}^{}_{l{\rm x}}\bar{\mathrm{A}}^{}_{m{\rm x}}\bar{\mathrm{A}}_{n{\rm y}} ^*     +   \hat{b}_{ml}^*\hat{b}_{pn}^*   \bar{\mathrm{A}}^{}_{l{\rm y}}\bar{\mathrm{A}}^{}_{m{\rm x}}\bar{\mathrm{A}}_{n{\rm y}} ^*       \nonumber \\
    &&~~\left.+ \hat{b}^{}_{lm}\hat{b}_{pn}^* \bar{\mathrm{A}}^{}_{l{\rm x}}\bar{\mathrm{A}}^{}_{m{\rm y}}\bar{\mathrm{A}}_{n{\rm y}} ^*    +  \hat{c}^{}_{lm}\hat{b}_{pn}^* \bar{\mathrm{A}}^{}_{l{\rm y}}\bar{\mathrm{A}}^{}_{m{\rm y}}\bar{\mathrm{A}}_{n{\rm y}} ^*         \right]{\Big)} \nonumber \\\label{eq_Npx}
\end{eqnarray}

We are interested in finding the ensemble average, or the expectation value, that we denote by $\langle~\rangle$ of the various coefficients appearing in (\ref{eq_Npx}) under random unitary transformations.  When the subscripts $l,m,p,n$ correspond to different spatial modes ($l\!\neq\! m\!\neq\! p\! \neq\! n$), the coefficient $a_{lm}a_{pn}$ is a sum of products of  functions $r_{11l},r_{11m},r_{11n},r_{11p}$ and $r_{21l},r_{21m},r_{21n},r_{21p}$, all of which are independent random variables with zero mean. As a result, $\langle a_{lm}a_{pn} \rangle=0$. With the same kind of reasoning, we find that all terms in (\ref{eq_Npx}) average to zero, except those containing the combinations
\begin{eqnarray}
    \langle \hat{b}_{pp}^{}\hat{b}_{pp}^* \rangle=1-\langle \hat{a}_{pp}^{}\hat{a}_{pp}^* \rangle&=&\frac{1}{3} \label{eq_s3}\\
    c_{mm}^{}a_{pp}^{} = c_{mm}^{}a_{pp}= a_{pp}^{}a_{pp}^{}= c_{pp}^{}a_{pp} &=&1 \label{eq_cmm}\\
    \langle \hat{a}_{pm}^{}\hat{a}_{pm}^* \rangle=\langle \hat{b}_{pm}^{}\hat{b}_{pm}^*
    \rangle &=&\frac{1}{2} \label{eq_amp}\\
    \langle a_{mp}^{}a_{pm}^{} \rangle=\langle b_{pm}^{}b_{pm}^* \rangle&=&\frac{1}{2} \label{eq_amp2}.
\end{eqnarray}
These equations are obtained using the properties of Haar matrices, which are given by Eq. (\ref{eq_RM1}) to (\ref{eq_RM4}) in the  case of $2M\times 2M$ matrices and assuming that $\mathbf{R}_i$ are independent random unitary matrices.
Finally, by gathering the nonzero terms together in $ \mathcal{N}_{p{\rm x}}$ and  $\mathcal{N}_{p{\rm y}}$ , we obtain the following averaged propagation equation:
\begin{eqnarray}
    \frac{\partial\bar{ \mathbf{A}}_p}{\partial z} &+& \langle{ \bm \delta}{\bm \beta}_{0p}\rangle \bar{\mathbf{A}}_p  + \langle{\bm  \delta}{\bm \beta}_{1p}\rangle
    \frac{\partial\bar{\mathbf{A}}_p}{\partial{t}} + \imath \frac{{\beta}_{2p}}{2}\frac{\partial^2\bar{\mathbf{A}}_p}{\partial t^2}\nonumber\\
    &=& \imath\gamma{\Big(} f_{pppp}\frac{8}{9}|\bar{\mathbf{A}}_{p}|^2+
    \sum_{m\neq p} f_{mmpp}\frac{4}{3}|\bar{\mathbf{A}}_{m}|^2{\Big)} \bar{\mathbf{A}}_{p},
\label{eq_Man}
\end{eqnarray}
with
\begin{eqnarray}
\langle {\bm\delta}{\bm \beta}_{0p}\rangle&=& \frac{1}{2}(\beta_{p{\rm x}}+\beta_{p{\rm y}})-\beta_g\\
\langle {\bm\delta}{\bm \beta}_{1p}\rangle&=& \frac{1}{2}(\frac{{\beta}_{p{\rm x}}}{\partial \omega}\Big|_{\omega_0} +\frac{{\beta}_{p{\rm y}}}{\partial \omega}\Big|_{\omega_0} )-\frac{1}{v_g}
\end{eqnarray}

These generalized Manakov equations are deterministic, as they contain no rapidly fluctuating random terms. Equations~\ref{eq_Man} represents an extension of the standard Manakov equations, for multimode fibers with birefringence. The first nonlinear term represents the \emph{intramodal} nonlinear effects, resulting from self-phase modulation (SPM), and it occurs for single-mode fibers as well with  the same coefficient 8/9. The second nonlinear term is new. Its origin lies in the \emph{intermodal} nonlinearities resulting from cross-phase modulation (XPM) among various fiber spatial modes. The averaging over birefringence fluctuations reduces the factor of 2 that is associated with XPM to a lower value of 4/3.
Note that the random quantity  $  \bar{\mathbf{{q}}}_{mp}$ averages to zero.

The new Manakov equations (\ref{eq_Man}) obtained for multimode fibers can be solved numerically much faster than (\ref{eq_main}), and it should help considerably in designing future SDM systems. However, before it can be used with confidence, we need to ensure that its predictions agree with those of (\ref{eq_main}). In the next two sections we verify this by considering a specific SDM system.

\section{Numerical simulations}  \label{sec_4}

In this section we compare the performance of a specific SDM system
for multimode fibers with and without random birefringence. Propagation equations are solved numerically with the split-step Fourier method.
The case with no birefringence  is obtained as a reference by solving (\ref{eq_NLSE}).
The case with birefringence is obtained  either by solving explicitly  (\ref{eq_main}) applying new random matrices at each step  \cite {Evang92} or by solving the new set of generalized Manakov Equations (\ref{eq_Man}). The comparison of the numerical results represents a test of the new Manakov equations.

For simplicity, we assume that each spatial mode carries a single-wavelength channel (no WDM) at a bit rate of 114~Gb/s. We make use of polarization multiplexing (PDM) together with the QPSK format so that the symbol rate is $R_s=28.5$ Gbaud. Each PDM-QPSK signal is constructed from an independent random bit stream that is Gray-mapped onto the QPSK symbols. The pulse shape corresponds to a raised-cosine spectrum with a roll-off factor of 0.2. Each PDM-QPSK signal is made of $2^{15}$ modulated symbols with $2^{17}$ samples per polarization. The injected power per spatial mode is $P_{\rm in}=7$~dBm.  This choice is deliberately high in order to observe large nonlinear impairments so as to represent a very stringent test of the new Manakov equations. .

The 1000-km transmission line consists of 10 sections of 100-km multimode fiber, each section being followed by an ideal erbium-doped fiber amplifier (EDFA) that can compensate for the span losses of 0.2~dB/km (assumed to be the same for each spatial mode).  At the end of the fiber link, we artificially degrade the signal by adding noise to account for the amplified spontaneous emission (ASE) added by all EDFAs. This is justified since the nonlinear interaction between ASE and the signal has been verified to be much smaller than the nonlinear interaction of the signal on itself.

\begin{table}
\caption{General shape of spatial mode distributions supported by the step-index and graded-index fibers used in numerical simulations.}
{\footnotesize \centering
\begin{tabular}{|c|c|c|}
\hline
Spatial distribution shapes & Step-index fiber & Graded-index fiber\\
\hline
\multirow{5}{*}{\includegraphics[width=0.18\columnwidth]{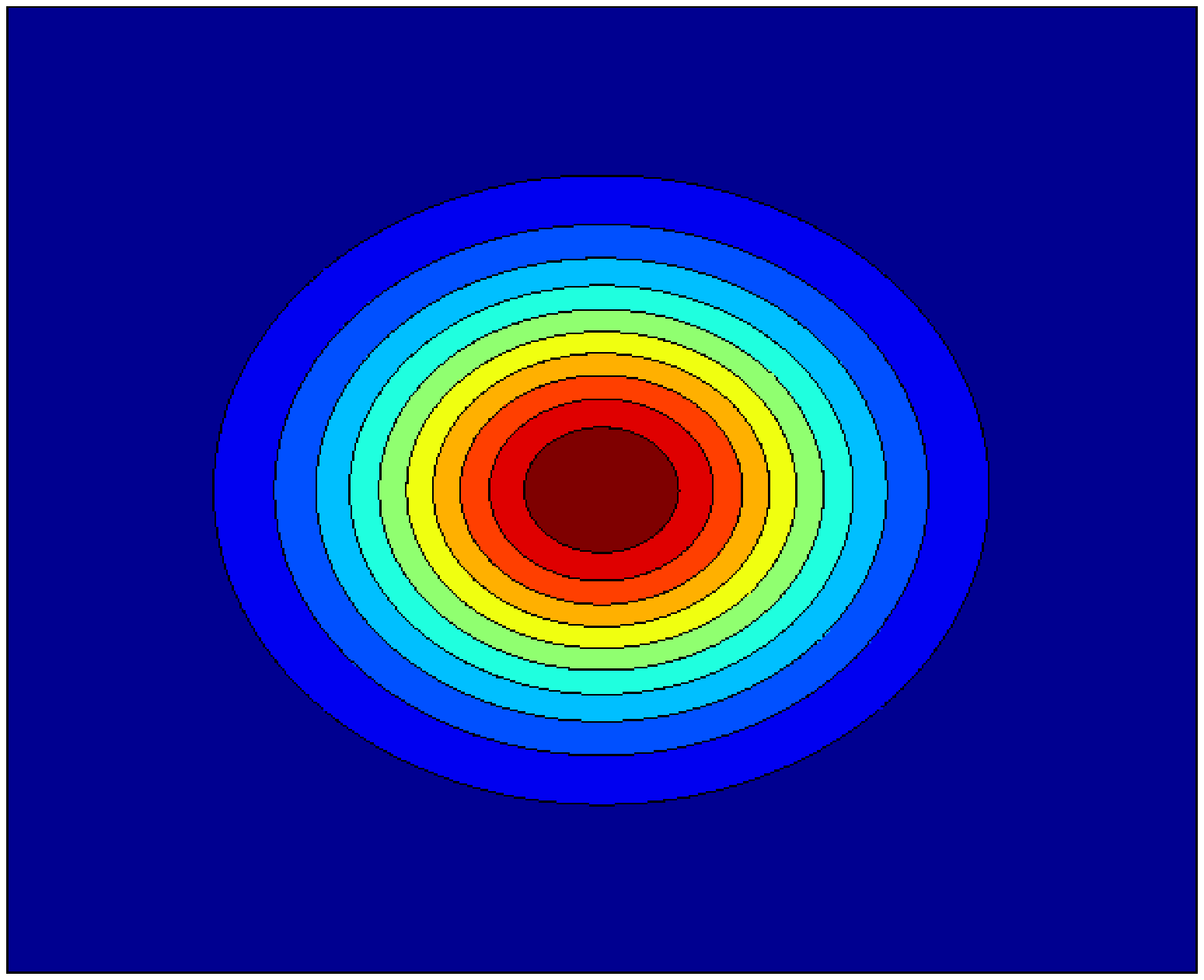} }&  & \\
 &  &  \\
 & LP01 & HG00 \\
 &  &  \\
 &  &  \\ \hline
\multirow{5}{*}{\includegraphics[width=0.18\columnwidth]{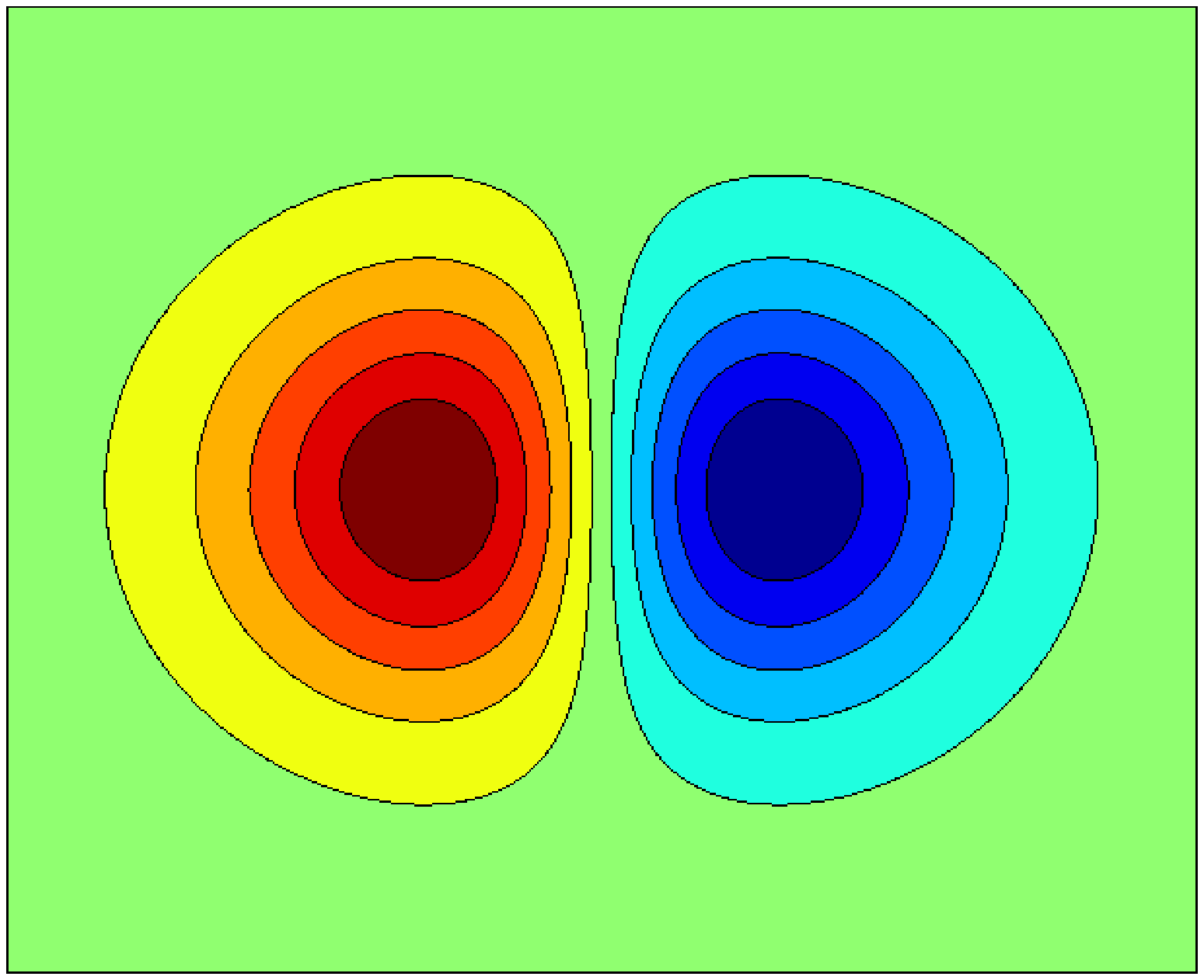}~\includegraphics[width=0.18\columnwidth]{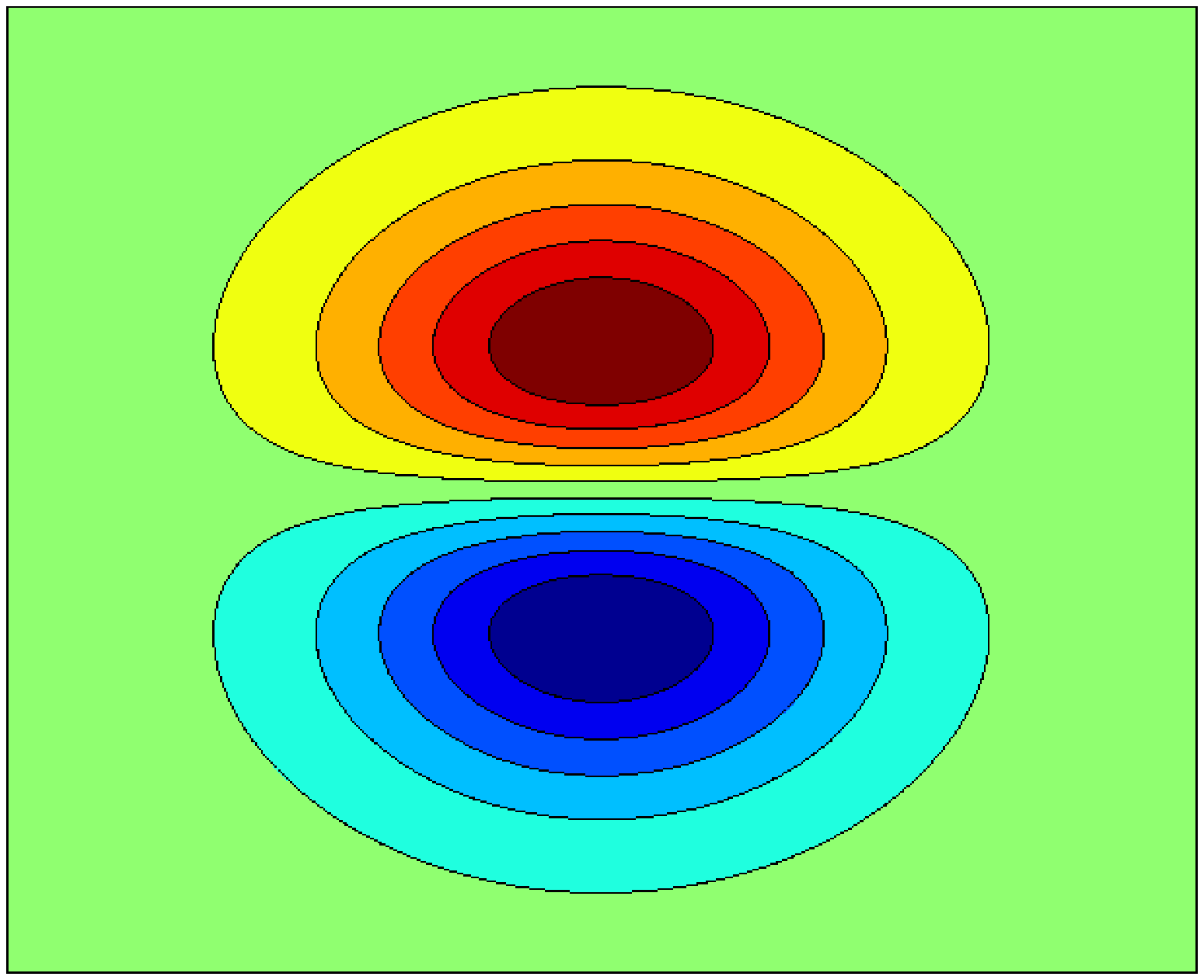}  }&  & \\
 &  &  \\
 & LP11a, LP11b & HG01, HG10 \\
 &  &  \\
 &  &  \\ \hline
\multirow{5}{*}{\includegraphics[width=0.18\columnwidth]{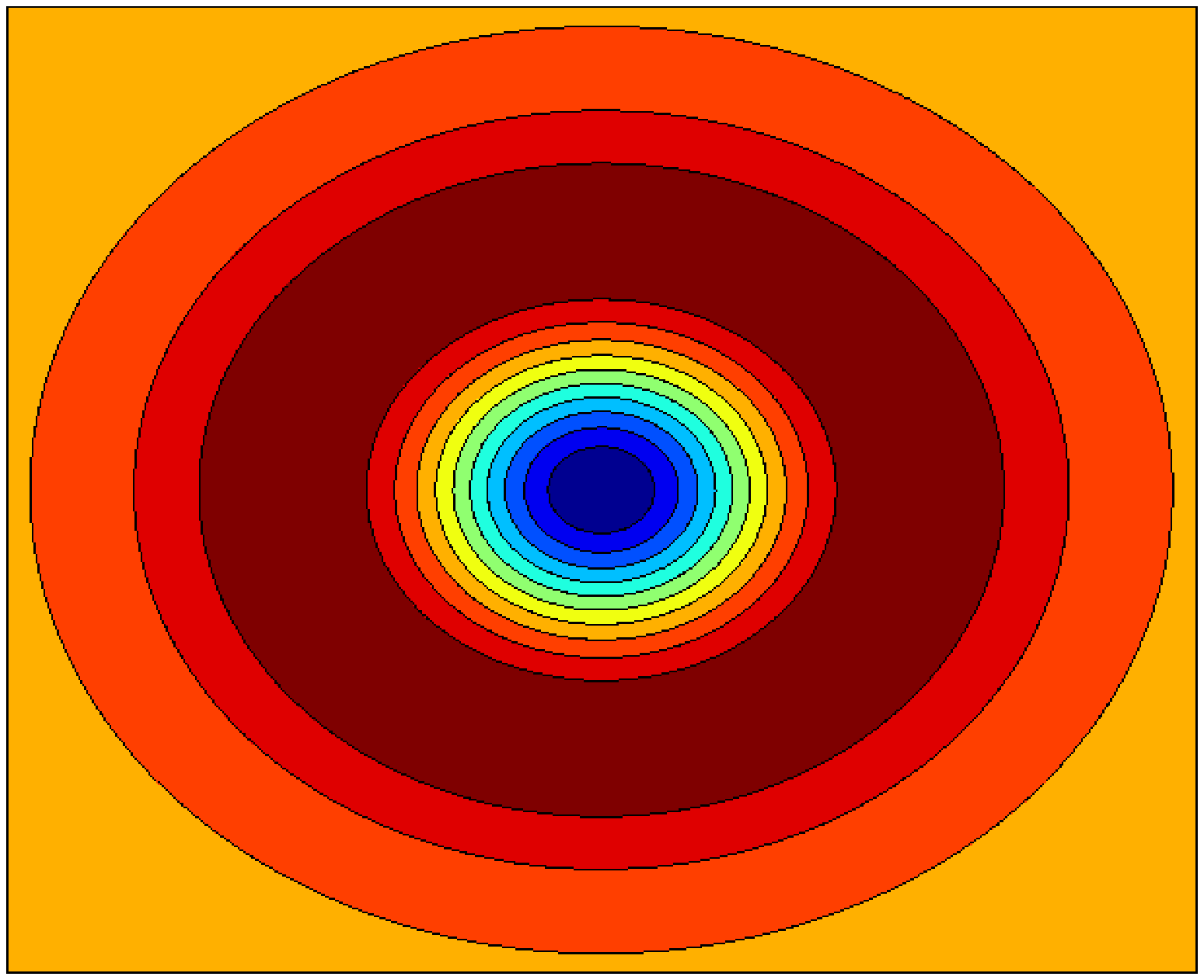} }&  & \\
 &  &  \\
 & LP02 & HG02+HG20 \\
 &  &  \\
 &  &  \\ \hline
 \multirow{5}{*}{\includegraphics[width=0.18\columnwidth]{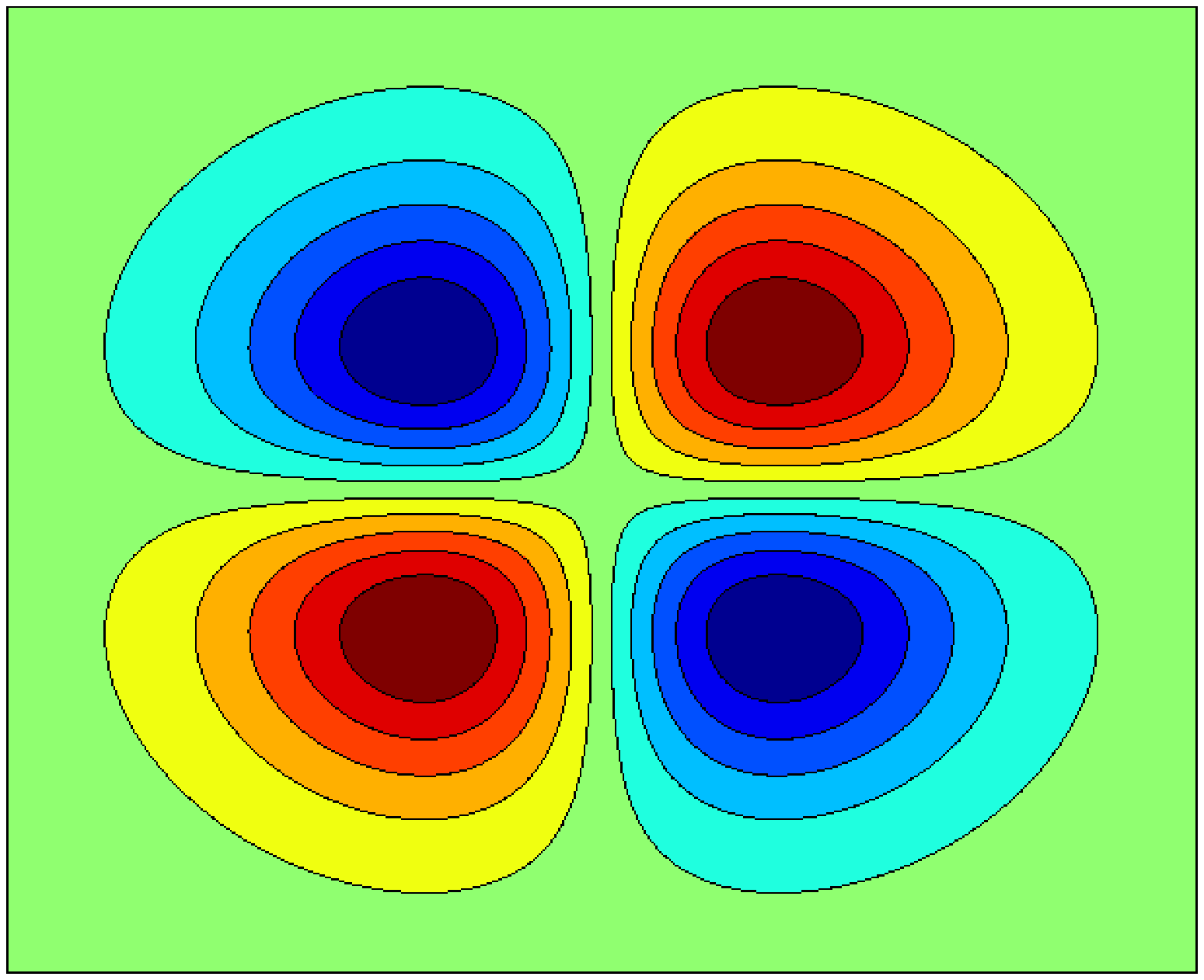}~\includegraphics[width=0.18\columnwidth]{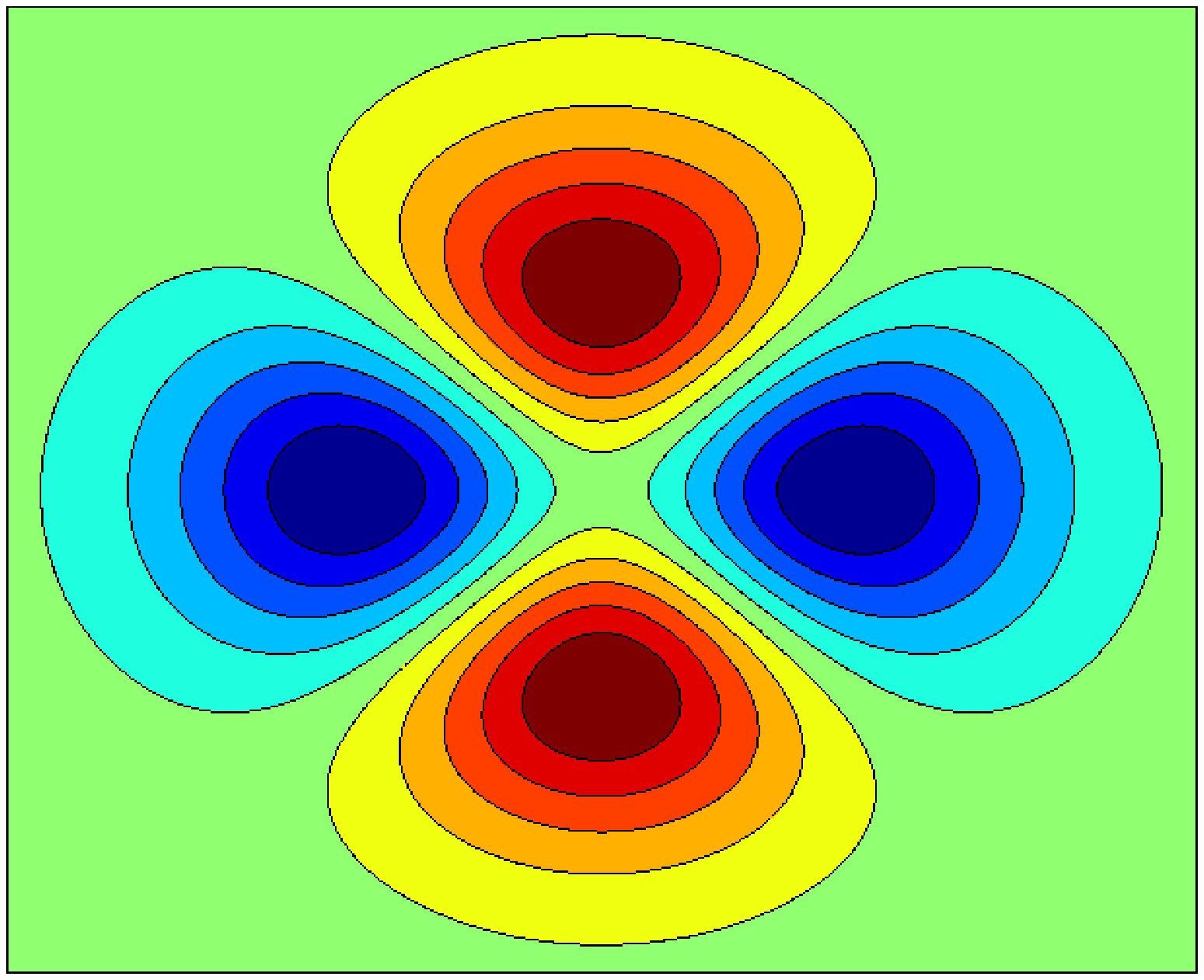}  }&  & \\
 &  &  \\
 & LP21a, LP21b & HG11a, HG11b \\
 &  &  \\
 &  &  \\ \hline
\end{tabular}
}
\label{tab_mode}
\end{table}

We focus on a step-index multimode fiber with a core diameter of $12.3~\mu$m and a numerical aperture of 0.2 ($\Delta=0.01$). Such a fiber has the V parameter of 5 at 1550 nm, and it supports  LP01, LP02, LP11, LP21 modes, resulting in a total of 6 spatial modes when we take into account 2-fold degeneracy of the LP11 and LP21 modes \cite{Ghatak}. Table~\ref{tab_mode} presents the spatial distributions $\mathrm{F}_m(x,y)$ of the different spatial modes supported by the fiber.

 Each spatial mode has its own propagation constant, modal group velocity and GVD parameter $D$ (see Tab.~\ref{tab_MM}). The fundamental-mode nonlinear coefficient is $\gamma=1.4\,$W$^{-1}$km$^{-1}$. We consider the case where there is no coupling between the spatial modes, i.e.  $\bar{\mathbf{q}}_{mp}=0$.

We solve numerically (\ref{eq_main}) and (\ref{eq_Man}) with a step size of 100~m in all cases (larger step sizes could have been used in the Manakov case). For simplicity, we choose $\delta\bar{\bm \beta}_{1p}=0$ in the case of (\ref{eq_main}).  Smaller step sizes were tested and shown to give the same results.  

At the receiving end, the symbol stream is coherently detected by  polarization diverse 90-degree hybrid followed by analog to digital converters. We assume an ideal DSP scheme that compensates perfectly all linear impairments such as chromatic dispersion, group-velocity mismatch, mode mixing, and polarization mixing. This is justified in practice since modern equalizers can remove most, if not all, linear impairments \cite{AlbertoPTL}. It also lets us present our results in a way that is independent of the type of equalizer employed.

\begin{table}
\caption{Differential modal group delay (compared to the fundamental mode) and group velocity dispersion $D$ of each spatial mode of the step-index multimode fiber used in section~\ref{sec_4}}
\center
\normalsize
\begin{tabular}{|c|c|c|}
  \hline
  &DMGD [ns/km] & $D$ [ps/(km-nm)] \\  \hline
  LP01&0 &25 \\
  LP11&6.5 & 27.3 \\
 LP02&9.9 & -2.3 \\
  LP21&12 & 20.8 \\
  \hline
\end{tabular}
\label{tab_MM}
\end{table}

\begin{figure}[t]
\begin{center}
\includegraphics[width=0.99\columnwidth]{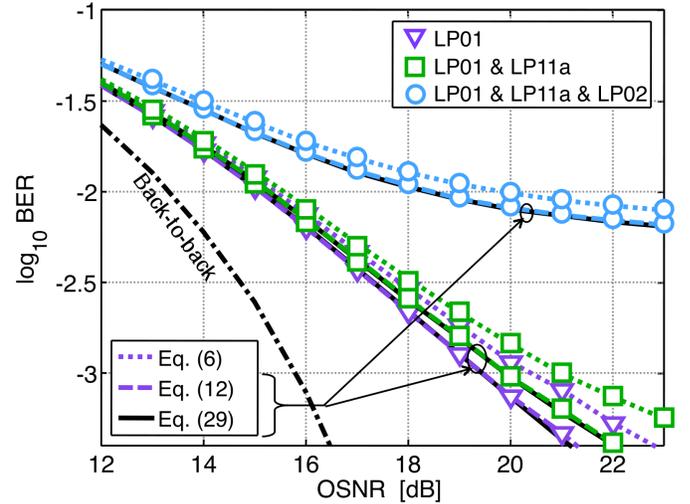}
\caption{\small  BER averaged over the co-propagating spatial modes  of a step index multimode fiber for   1000~km transmission versus OSNR.  114-Gb/s PDM-QPSK signals are transmitted over different combinations of  spatial modes indicated in the upper right inset.  Solid curves obtained  solving Manakov Eq.~(\ref{eq_Man}) are superposed with dashed curves representing the case with random birefringence  obtained solving (\ref{eq_main});  pointed curves represent the zero-birefringence case (\ref{eq_NLSE}). Back-to-back case (no fiber) is displayed as a reference.}
\label{fig_MMmanakov}
\vspace{-6mm}
\end{center}
\end{figure}

\begin{figure}[t]
\begin{center}
\includegraphics[width=0.99\columnwidth]{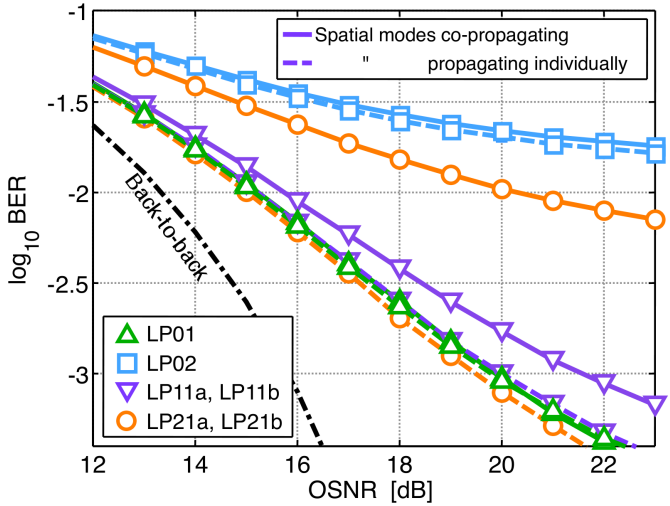}
\caption{\small BER of each spatial mode versus OSNR  after  1000~km  transmission on step-index fiber with rapidly varying birefringence. Solid curves represent the case where 6 spatial modes co-propagate; dashed curves represent the cases where spatial modes propagate individually.}
\vspace{-6mm}
\label{fig_MMV5allmode}
\end{center}
\end{figure}

Figure~\ref{fig_MMmanakov} shows the bit-error rate (BER) averaged over all the propagating spatial modes as a function of the optical signal-to-noise ratio (OSNR) \cite{Essiambre_JLT} after 1000~km multimode fiber when 114-Gb/s bit streams are transmitted on different combinations of spatial modes.  Clearly, the predictions of our new Manakov equation~(\ref{eq_Man}) agree very well with those of the full stochastic equation (\ref{eq_main}). As discussed later, the use of Manakov equations reduces considerably the computation time.
To study the impact of random birefringence on system performance, we also consider the case of an ideal multimode fiber with no birefringence by solving Eq.~(\ref{eq_NLSE}). 
It is evident that birefringence actually improves system performance in all cases by lowering the BER. It should be stressed that this performance improvement occurs in practice only if the random polarization mixing resulting from birefringence can be efficiently compensated at the receiver. 
The system performance is easily understood if we recall that the averaging over rapidly varying birefringence leads to (\ref{eq_Man}) in which the nonlinear coefficient associated with the SPM is reduced from 1 to 8/9. Note that for the chosen configurations of transmitted spatial modes, inter-modal XPM effects are averaged out because of the large differential group delay between spatial modes \cite{Koeb11}. 
The reason of the large performance degradation observed in Fig.~\ref{fig_MMmanakov} in the case of 3 co-propagating spatial modes (LP01+LP11+LP02)   is related to a small magnitude of the dispersion parameter $D$ of the LP02 spatial mode for the chosen fiber design parameters. Figure~\ref{fig_MMV5allmode} presents the received BER of each spatial mode. In one case, the 6 spatial modes are transmitted together. In the other case, each spatial mode is transmitted individually in the multimode fiber. The performance degradation between these two cases is due to the inter-modal XPM. We observe that LP02 and LP01 spatial mode suffer very little from inter-modal XPM because of their large differential group delay compared to the other spatial modes. On the other hand, performance of both LP11 and LP21 modes are largely degraded by the inter-modal XPM because the degenerated spatial modes (for instance LP11a and LP11b) propagate at the same group velocity. Note that two degenerated spatial mode perform the same. The nonlinear inter-modal XPM coefficient is $f_{11_a,11_a,11_b,11_b}=0.3$ between LP11a and LP11b and  $f_{21_a,21_a,21_b,21_b}=1$ between LP21a and LP21b.

In a step-index multimode fiber,  there is a spread of group velocities between spatial modes. This group-velocity mismatch introduces linear degradations (similar to PMD) that must be compensated at the receiver. In practice, the equalization complexity grows in proportion to the magnitude of the differential group delays among  modes. This problem can be solved using a graded-index fiber for which group-velocity mismatch nearly vanishes. It is thus important to study how the system performance in Fig.~\ref{fig_MMmanakov} will change if the step-index multimode fiber used there were replaced with a graded-index multimode fiber. In particular, we focus on a multimode fiber whose core has a quadratic graded refractive index (in the shape of a parabola). Such fibers are designed to have all spatial modes propagate at the same group velocity~\cite{Ghatak}.

\begin{figure}[t]
\begin{center}
\includegraphics[width=0.99\columnwidth]{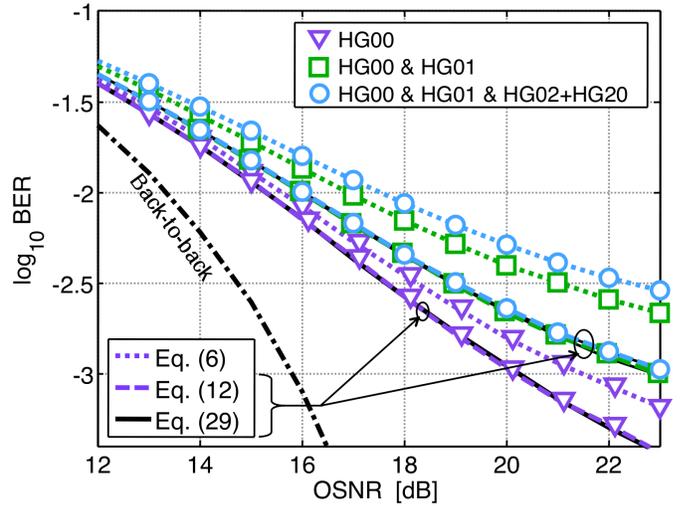}
\caption{\small BER averaged over the co-propagating spatial modes  of a step index multimode fiber for   1000~km transmission versus OSNR.  114-Gb/s PDM-QPSK signals are transmitted over different combinations of  spatial modes indicated in the upper right inset.  Solid curves obtained solving Manakov Eq.~(\ref{eq_Man}) are superposed with dashed curves representing the case with random birefringence  obtained solving (\ref{eq_main});  pointed curves represent the zero-birefringence case (\ref{eq_NLSE}). Back-to-back case (no fiber) is displayed as a reference.}
\vspace{-6mm}
\label{fig_MMGImanakov}
\end{center}
\end{figure}

For numerical simulations, we consider a graded-index fiber with a core diameter of 17.4~$\mu$m, $D=21.5$ ps/(km-nm) and  $\gamma=1.4$ W$^{-1}$/km and designed such that it supports the (Hermite-Gauss) modes presented in Table~\ref{tab_mode}. Data transmitted on different combinations of these spatial modes and we assume no linear coupling between them, i.e.  $\bar{\mathbf{q}}_{mp}=0$. For similarity with LP02 spatial mode of step-index fibers, we consider a spatial distribution corresponding to the sum of  the HG02 and HG20 spatial mode distributions.
Figure \ref{fig_MMGImanakov} shows the BER as a function of OSNR under the same conditions used in Fig.~\ref{fig_MMmanakov} except that the graded-index fiber replaces the step-index fiber. A comparison of two figures shows several interesting features. First, a rapidly varying birefringence results in a larger improvement in the case of graded-index fibers than in the case of step-index fibers. Here, spatial modes have the same group velocity so, there are nonlinear interactions between them due to  inter-modal XPM. The effect of rapidly varying birefringence results to reduction of the XPM coefficient from 2 to 4/3, which improves the performance.
Second, nonlinear penalties, for graded-index fibers, are larger in the two-mode case but become considerably smaller in the three-mode case. These results suggest that graded-index multimode fibers may perform better when a large number of spatial modes are used to simultaneously transmit data.

Finally, we come to a main advantage of using the new  Manakov equations (\ref{eq_Man}). We have already seen that its numerical predictions are virtually identical to those obtained by solving Eq.~(\ref{eq_main}). From a practical standpoint, the use of Manakov equations is preferable because it requires much less computational time. One reason is that the number of nonlinear terms is drastically reduced. For instance in the case of 3 propagating modes, there are 81 nonlinear terms in  Eq.~(\ref{eq_main}) but only 3 terms are needed in Eq.~(\ref{eq_Man}). In general, there are $M^3-1$  intermodal nonlinear terms for a $M$-mode fiber while there are $M-1$ intermodal nonlinear terms for each generalized Manakov equation. A second reason is that a solution of Eq.~(\ref{eq_main})  requires  computation of an exponential matrix of size $2M\times2M$, $M$ being the number of spatial modes, which is a time-consuming numerical operation. As a result, the reduction in computing time depends on the value of $M$. This is evident in Table~\ref{tab_1}, which shows the computation time obtained on a desktop computer using the MATLAB software (version 2007b)  for 1000-km propagation of 114-Gb/s channels with a step size of 100\,m and a temporal grid size of $2^{17}$. Computation time is reduced by nearly a factor of 10 even for $M=2$, with larger reductions occurring for larger values of $M$.  Identical step sizes were used in both cases.
Solving Eq.~(\ref{eq_main}) numerically requires a step size that is determined by  birefringence fluctuations. In the case of Manakov equations~(\ref{eq_Man}), the step size is mainly set by nonlinear effects which usually occur on a length scale much larger than that of birefringence fluctuations. Consequently it is usually possible to choose a much larger step size. Taking this into account, the use of the new Manakov equations should reduce the computing time easily by more than a factor of 100 in most cases of practical interest.

 \begin{table}[tb!]
\caption{\small Computing time in seconds for 1000-km transmission of 114-GB/s channels over multiple modes with a 100-m step size.}
\begin{center}
\normalsize
\begin{tabular}{|c|c|c|c|c|}
\hline
Number of modes, $M$ & 1 & 2 & 3 &4\\
\hline
Manakov Eq.~(\ref{eq_Man}) & 270 & 800 & 1600& 2800 \\ \hline
unaveraged Eq.~(\ref{eq_main}) & 720& 7600& 16000& 31000 \\\hline
\end{tabular}
\end{center}
\label{tab_1}
\end{table}%

\section{Multicore Fibers}   \label{sec_5}

Manakov equations~(\ref{eq_Man}) in Section \ref{sec_3} assume weak coupling among distinct spatial modes of a multimode fiber. In practice, linear mode coupling can stem from such effects as core irregularities \cite{Marcuse} or fiber bending \cite{Shem09}.  Linear coupling between distinct spatial modes can be relatively weak in comparison to the coupling between the two polarization components of each mode.  In our theory, linear mode coupling resulting from fiber imperfections is governed by the parameter $q_{mp}(z)$ appearing in Eq.~(\ref{eq_NLSE}). This coupling coefficient varies randomly along the propagation distance and its strength depends on the degree of fiber imperfections and the level of stress applied to the fiber.

Future SDM systems may also make use of multicore fibers in which each core supports a single mode, but modes in different cores can couple strongly if these cores are relatively close to each other. In this case, the extent of coupling depends on the physical distance among its cores and its magnitude varies exponentially with this distance. The coupling coefficient $q_{mp}(z)$ may vary with $z$ in practice due to waveguide imperfections but in order to simplify the analysis, we assume ideal fiber so that $q_{mp}$ is a constant in the multicore study.

For numerical simulations, we consider an ideal 3-core fiber having identical, and equidistant single-mode cores, i.e., in an equilateral triangular configuration. There are three spatial modes $\mathrm{F}_m$ each associated with the field propagating in one core. The nonlinear propagation equation is still given by (\ref{eq_NLSE}) but the coupling coefficients are now given by 
\begin{eqnarray}
     {q}_{mp}&=&\frac{k_0^2}{2{\bm \beta}_{0p}({\rm I}_m{\rm I}_p)^{1/2}}\iint (n^2-n_m^2)\mathrm{F}_m^{}\mathrm{F}_p\,dx\,dy,
\end{eqnarray}
where $n_m(x,y)$ is the spatial distribution of the refractive index of the isolated cores  in which the spatial mode $m$ propagates. This propagation equation is obtained using a coupled-core approach, which assumes that the spatial distribution of a field propagating in one core, is not perturbed by the presence of other cores \cite{Marcuse}. Because of the symmetry of the three-core fiber considered here,  all coupling coefficients are identical and we can replace ${q}_{mp}$ with a single parameter $q$. Each core of the fiber has a core diameter of 8.2$\,\mu$m, with $\Delta=0.003$ and $\gamma=1.4$ W$^{-1}$km$^{-1}$. With these design parameters, intermodal nonlinear coefficients are negligible \cite{Mumt12_PTL}. We modify the strength of linear coupling by adjusting the distance between fiber cores. It is useful to present the results in terms of the coupling length $L_c=\pi/(3q)$, which represents the propagation distance  at which maximum linear power transfer occurs among the cores. The rapidly varying birefringence is included by solving the propagation Eq.~(\ref{eq_main}) with a step size of 100~m and changing the rotation matrix randomly after each step.

\begin{figure}[tb!]
\begin{center}
\includegraphics[width=0.99\columnwidth]{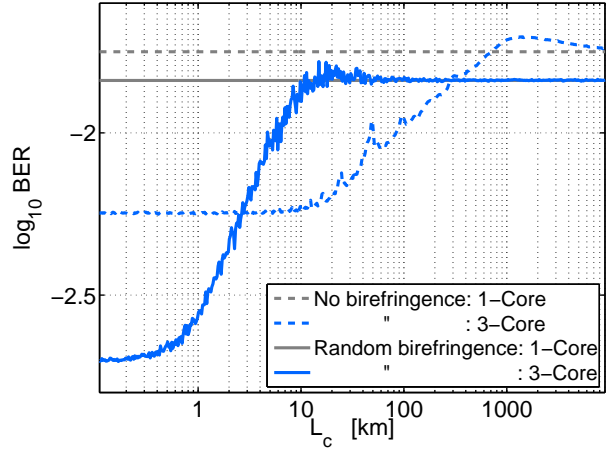}
\caption{\small BER averaged over the cores as a function of coupling length $L_c$ for 1000~km transmission. 114-Gb/s PDM-QPSK with independent random bit streams are transmitted over a single-core (solid horizontal line) and a 3-core (solid curve) fiber. Dashed curves show these two cases for fibers without birefringence.}
\vspace{-6mm}
\label{fig_MCmanakov}
\end{center}
\end{figure}

Figure \ref{fig_MCmanakov} shows the BER as a function of the coupling length (solid blue curve) when 114-Gb/s independent random bit streams are simultaneously transmitted through all three cores. The case of a single-core fiber is shown for comparison by a solid black line. Also shown for comparison are the two dashed curves for which the fiber link has no birefringence. The injected power per core is $P_{\rm in}= 8$~dBm and the OSNR at the receiver is set to $20$~dB in all cases.
When the coupling length is comparable to the length scale $L_f$ of birefringence fluctuations (here 0.1~km), there is a large performance improvement in the case of the 3-core fiber. This improvement can be understood by noting that a rapid power transfer among various modes induced by a strong linear coupling and birefringence-induced polarization mixing helps to mitigate nonlinear effects. As predicted by the new Manakov equations (\ref{eq_Man}) (where the term $\bar{\mathbf{q}}_{mp}$ in~(\ref{eq_17})  averages to zero), when the coupling length becomes much larger than $L_f$,  the effects of linear coupling are averaged out by rapidly varying birefringence, and the 3-core fiber behaves similarly to a single-core fiber. This occurs in Figure \ref{fig_MCmanakov} for $L_c>10$~km. Indeed, when Eq.~(\ref{eq_Man}) is used,  the numerical results correspond to the case of a single-core fiber. The main message from Figure~\ref{fig_MCmanakov} is that  the new Manakov equations (\ref{eq_Man}) can be used with success for multicore fibers as long as the cores are far enough apart that the coupling length is about 100 times larger than the length scale of birefringence fluctuations.

These results also suggest that in the case of multimode fibers, Manakov equations (\ref{eq_Man}) are valid when the coupling between spatial modes, resulting from fiber imperfections, occurs on a length scale 100 times larger than that of  birefringence fluctuations.

\section{Manakov Equations in strong-Coupling Regime} \label{sec_6}

The Manakov equations (\ref{eq_Man}) obtained in the weak-coupling regime fail if linear coupling among various spatial modes becomes comparable to  birefringence-induced coupling. In this section, we derive new Manakov equations in the case of strong random coupling among $M$ spatial modes of a multimode or multicore fiber. In this strong-coupling regime, we need to use $2M\times 2M$ random unitary matrices to account for random coupling among $M$ spatial modes, with two polarization states per mode.

Before applying such matrices to the nonlinear propagation equation, we need to rewrite the set (\ref{eq_NLSE}) of $M$ equations ($p=1\ldots M$) in a vector form. By introducing $\mathcal{A}=[\mathbf{A}_1^{\rm T}\ldots \mathbf{A}_{M}^{\rm T}]^{\rm T}$ as the column vector containing $2M$ field envelopes, we obtain
 \begin{eqnarray}
    \frac{\partial\mathcal{A}}{\partial{z}}&=&\imath \delta{\mathcal{B}}_{0} \mathcal{A}-{\mathcal{B}}_{1} \frac{\partial\mathcal{A}}{\partial{t}}  - \imath\frac{{\mathcal{B}}_{2}}{2}\frac{\partial^2\mathcal{A}}{\partial{ t}^2} \nonumber\\
    &+&\iint\frac{\imath\gamma}{3} \left[
 (\mathcal{A}\mathcal{F}\mathcal{A})\mathcal{F}^*\mathcal{A}^*\right.\left.+2
 (\mathcal{A}^{\rm H}\mathcal{F}\mathcal{A})\mathcal{F}\mathcal{A}\right] dx\,dy\nonumber\\\label{eq_NLSE2}
\end{eqnarray}
where $\mathcal{F}$ is a $2M\times 2M$ matrix whose diagonal elements are $M$ diagonal $2\times 2$ matrices such that 
 \begin{eqnarray}
    \mathcal{F}_{ij}&=&\mathrm{F}_i\,\mathrm{F}_j\,\mathbf{I}_2, \qquad\qquad i,j=1\ldots M.
 \end{eqnarray}
${\mathcal{B}}_{0} $, ${\mathcal{B}}_{1} $ and ${\mathcal{B}}_{2} $ are  $2M\times 2M$ diagonal matrices containing respectively the propagation constant, the inverse group velocity, and the dispersion parameter of each mode. Further,
\begin{equation}
    \delta{\mathcal{B}}_{0}= \mathcal{B}_0- \frac{1}{2M}{\rm Tr}\left( \mathcal{B}_0\right)
\end{equation}
represents deviations of the propagation constants from the average value of all spatial modes. Note that the coupling term $q_{mp}$ does not appear in (\ref{eq_NLSE2}) as it is included through the random unitary matrix $\mathcal{R}$.

Following the procedure in section~\ref{sec_3}, we apply the substitution
\begin{equation}
    \mathcal{A}=\mathcal{R}\bar{\mathcal{A}}~, \label{eq_HCsub}
\end{equation}
where $\mathcal{R}$ is a random rotation matrix of dimension $2M\times 2M$  whose elements $r_{ij}$ satisfy the following relations~\cite{Antonia}:
\begin{eqnarray}
    \langle|r_{ik}|^2|r_{ik'}|^2\rangle&=&\frac{1+\delta_{kk'}}{2M(2M+1)}
\label{eq_RM1}\\
    \langle|r_{km}|^2|r_{k'm'}|^2\rangle&=&\frac{1}{(2M+1)(2M-1)}\quad~~
    \begin{array}{c}\\k\neq k'\\m\neq m'\end{array}\\
    \langle r_{ki}^*r^{}_{kj}r^{}_{k'i}r_{k'j}^*\rangle &=&\frac{-1}{(2M+1)2M(2M-1)}\quad~
    \begin{array}{c}\\k\neq k'\\i\neq j\label{eq_RM4}\end{array}
\end{eqnarray}

By applying the substitution (\ref{eq_HCsub}) and averaging the propagation equation ~(\ref{eq_NLSE2}), we obtain the following generalization of the Manakov equations in the strong-coupling regime:
 \begin{eqnarray}
    \frac{\partial\mathcal{A}}{\partial{z}}&+&\frac{1}{v}\frac{\partial\mathcal{A}}
    {\partial{t}} + \imath \frac{\bar{\beta}_{2}}{2}\frac{\partial^2\mathcal{A}}{\partial{ t}^2}
    =~~\imath \gamma\kappa |\mathcal{A}|^2\mathcal{A} \label{eq_ManaHC}
\end{eqnarray}
where
\begin{equation}
    \kappa= \sum^{M}_{k\le l}\frac{32}{2^{\delta_{kl}}}
    \frac{{f}_{kkll}}{6M(2M+1)}~,\label{eq_kap}
\end{equation}
$1/v={\rm trace}\left( \mathcal{B}_1\right)/2M$ is the average inverse group velocity, and $\bar{\beta}_{2}={\rm trace}\left( \mathcal{B}_2\right)/2M$ is the average group velocity dispersion. Equations (\ref{eq_ManaHC}) and (\ref{eq_kap}) are in agreement with the results obtained in \cite{Meco12} using a different procedure.

\begin{figure}[t]
\begin{center}
\includegraphics[width=0.99\columnwidth]{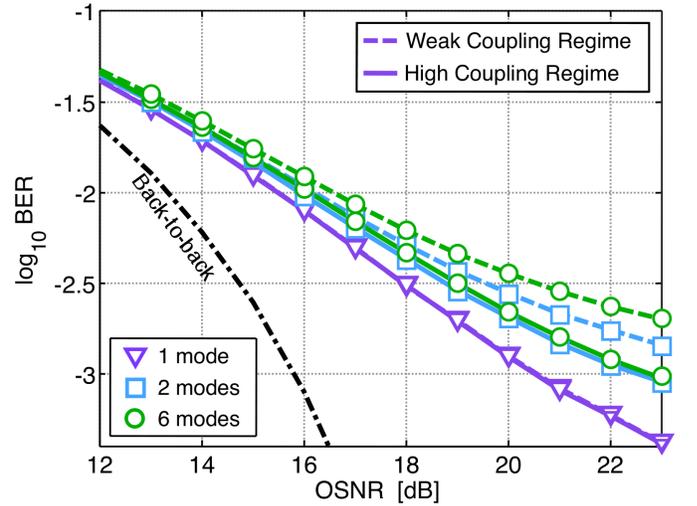}
\caption{\small BER averaged over the co-propagating modes versus OSNR after 1000~km over  graded-index fibers. 114-Gb/s bit  PDM-QPSK signals are transmitted on one, three and six spatial modes. Square and diamond markers represent the weakly and strong coupling regimes, respectively. The dotted curve shows the back-to-back case (no fiber).}
\vspace{-6mm}
\label{fig_WCVsHC}
\end{center}
\end{figure}

A comparison of the magnitudes of various nonlinear coefficients appearing in~(\ref{eq_Man}) and~(\ref{eq_ManaHC}) reveals that  system performance are improved in the high-coupling regime as the number of spatial modes increases. The nonlinear coefficient $\kappa$ in Eq.~(\ref{eq_kap}) tends to zero with $M$ increasing, which predicts that multimode fibers in the high coupling regime can perform better than single-mode fibers for large number of propagating spatial modes. 
Using the same parameter values used in section \ref{sec_4},  Figure~\ref{fig_WCVsHC} presents  numerical results for 1000-km transmission over a graded-index fibers  supporting one, three and seven Hermite-Gaussian HG$_{mn}$ spatial modes. Each fiber is designed in order to have $\gamma=1.4$ W$^{-1}$/km. With the three mode fiber,  data are transmitted on the HG00, HG01 and HG10 spatial modes. With the seven mode fiber,  data are transmitted on the six spatial modes presented in Table.~\ref{tab_mode}. 
Comparison of the weak and strong coupling regime shows that, as expected, the BER curves in the two cases coincide when only one spatial mode of the fiber is used. However, the high-coupling regime results in better BER performance when data is transmitted using multiple spatial modes. Another difference is that the performance is degraded in the weak-coupling regime as the number of spatial modes increases whereas, somewhat surprisingly, performance is nearly the same for three or six spatial modes in the high-coupling regime.

\section{Conclusions}

In this paper we have focused on SDM systems designed by using multimode or multicore fibers and derived a nonlinear propagation equation satisfied by the bit stream transmitted through each optical mode in the presence of fiber dispersion, birefringence, and nonlinearity. We first expressed this equation in  Jones vector form in the slowly varying envelope approximation so that both polarization components of each mode can be treated simultaneously and we then used it to investigate the performance of SDM systems designed with multimode fibers exhibiting modal birefringence that varies randomly on a length scale smaller than the length over which nonlinear effects become important. The effects of fluctuating birefringence were included through random Jones matrices. As expected, the XPM and FWM effects resulting from intermodal nonlinearities cannot be ignored for multimode fibers. There are as many as $M^3-1$ intermodal nonlinear terms for an $M$-mode fiber.

Since numerical simulations based on such stochastic nonlinear propagation equations are generally time-consuming, we generalized the Manakov equations, well known in the case of single-mode fibers, to the case of multimode fibers. We considered the weak linear coupling case first and averaged over birefringence fluctuations following a procedure similar to that used in the original derivation of the Manakov equations for single-mode fibers. We found that averaging over birefringence fluctuations reduces the $M^3-1$ intermodal nonlinear terms to only $M-1$ terms because all FWM-type terms disappear. The XPM-type terms remain but their effectiveness is reduced because the well-known factor of 2 is reduced to 4/3.

We verified the validity of our new Manakov equations by simultaneously transmitting multiple 114-Gb/s bit streams in the PDM-QPSK format over different modes of a multimode fiber and comparing the numerical results with those obtained by solving the full stochastic equation explicitly. The agreement between the two methods was excellent in all cases studied. We found that the use of the new Manakov equations can reduce computation time by at least a factor of 10 and by a factor in excess of 100 as the number of modes grows. Our numerical results show that birefringence fluctuations improve system performance by reducing the impact of fiber nonlinearities. The extent of improvement depends on the fiber design and how many modes are used for SDM transmission. We investigated both the step-index and graded-index multimode fibers and considered up to three spatial modes in each case.

We extended our theory to the case of multicore fibers and discussed the validity of the new Manakov equations by studying the impact of linear coupling among various cores. We found that the new Manakov equations remain valid when the length scale of random birefringence of the cores is about 100 times smaller than the linear coupling length among the cores of a multicore fiber.

Finally we studied the case of strong random coupling among various spatial modes of a multimode fiber and derive new Manakov equations for this regime. We showed though numerical simulations that the high-coupling regime leads to better system performance than the weak-coupling regime when the number of spatial modes is at least two. Theory also predicts that performance improves for large number of propagating spatial modes so that such multimode fibers can perform even  better than single-mode fiber.


\end{document}